\documentclass[letter]{aa} 

\usepackage{graphicx}
\usepackage{txfonts}
\usepackage{graphicx}
\usepackage{nth}
\usepackage{longtable}
\usepackage{color}
\usepackage{hyperref}
\usepackage{txfonts}
\usepackage{natbib}
\bibpunct{(}{)}{,}{a}{}{,}

\usepackage[caption=false]{subfig}

\defcitealias{Crossfield17}{C17}
\defcitealias{Rodriguez17}{R17}

\begin{document}

   \title{Ground-based photometry of the 21-day Neptune HD106315c$^{,}$\thanks{The photometric time series data are only available in electronic form
at the CDS via anonymous ftp to cdsarc.u-strasbg.fr (130.79.128.5) or via http://cdsweb.u-strasbg.fr/cgi-bin/qcat?J/A+A/}}


   \author{M. Lendl
          \inst{1,2}
           \and
           D. Ehrenreich
           \inst{2}
           \and
           O.D.Turner
           \inst{2}
           \and 
           D. Bayliss
           \inst{2}
           \and
           S. Blanco-Cuaresma
           \inst{2,3}
           \and
                   H. Giles
           \inst{2}
           \and
           F. Bouchy
           \inst{2}
           \and
           M. Marmier
           \inst{2}
           \and 
           S. Udry
           \inst{2}
}

   \institute{Space Research Institute, Austrian Academy of Sciences, Schmiedlstr. 6, 8042 Graz, Austria\\
              \email{monika.lendl@oeaw.ac.at}
              \and
              Observatoire de Gen\`eve, Universit\'e de Gen\`eve, Chemin des maillettes 51, 1290 Sauverny, Switzerland
              \and
              Harvard-Smithsonian Center for Astrophysics, 60 Garden Street, Cambridge, MA 02138, USA
         }

   \date{}

  \abstract
  {Space-based transit surveys such as \emph{K2} and the \emph{Transiting Exoplanets Survey Satellite (TESS)} allow the detection of small transiting planets with orbital periods greater than 10~days. 
  Few of these {warm} Neptunes are currently known around stars bright enough to allow for detailed follow-up observations dedicated to their atmospheric characterization. 
  The 21-day period and 3.95 ~$R_\oplus$ planet HD~106315c has been discovered by \emph{K2} based on the observation of two of its transits. 
  We observed HD~106315 using the 1.2~m Euler telescope equipped with the EulerCam camera on two occasions to confirm the transit using broadband 
  photometry and refine the planetary period. Based on two observed transits of HD~106315c, we detect its $\sim$1~mmag transit and obtain a precise measurement of 
  the planetary ephemerides, which are critical for planning further follow-up observations. We  used the attained precision together with the predicted yield from the 
  \emph{TESS} mission to evaluate the potential for ground-based confirmation of Neptune-sized planets found by \emph{TESS}. We find that one-meter class telescopes on 
  the ground equipped with precise photometers could substantially contribute to the follow-up of 162 \emph{TESS} candidates orbiting stars with magnitudes of $V \leq 14$. 
  Of these candidates, 74 planets orbit stars with $V \leq 12$ and 12 planets orbit $V \leq 10$, which makes them high-priority objects for atmospheric characterization with high-end instrumentation.}

   \keywords{planetary systems -- stars: individual: HD106315 -- techniques: photometric}

   \maketitle
%

\section{Introduction}
Since the repurposing of the \emph{Kepler} satellite \citep{Borucky09}, the \emph{K2} mission \citep{Howell14} has been surveying a set of fields along the ecliptic 
for transiting planets. Largely increasing the number of bright stars observed compared to \emph{Kepler's} original mission, \emph{K2} has been discovering an increasing 
number of small transiting planets orbiting bright stars  \citep[e.g.,][]{Vanderburg15,Vanderburg16a,Armstrong15,Crossfield15a,Petigura15}. %
Thanks to their bright hosts, these objects are prime targets for atmospheric studies through optical and near-IR transmission spectroscopy 
both from the ground \citep[e.g.][]{Bean10,Redfield08,Lendl16,Wyttenbach15} and from space \citep{Charbonneau02,Deming13,Sing15}, as well as 
exospheric characterization through UV observations \citep[e.g.,][]{Vidal03,Fossati10,Ehrenreich15}. Radial velocity observations efficiently provide precise 
planetary masses, and with well-determined stellar properties these objects are key for determining planetary mass-radius relations. After \emph{K2}, the \emph{Transiting Exoplanets Survey Satellite} \citep[\emph{TESS},][]{Ricker15}, 
foreseen for launch in 2018, will perform a nearly all-sky survey of bright stars, surveying 26 fields for at least 30~days each.

Of particular interest are planets at periods longer than $\sim$10 days, which are outside the detection realm of ground-based surveys such as the \emph{Next-Generation Transit Survey (NGTS)}
\citep{Wheatley13}, and inhabit a position in the parameter space currently ill-populated by objects bright enough for detailed follow-up. \footnote{A query of exoplanet.eu and exoplanets.org on 21 May 2017 reveals only 12 planets with $V<10$~mag, $R_P<0.5 R_J$ and $P>15$~days} One such object is the V=8.95 F5V star HD106315, which was observed by \emph{K2} during campaign 10, and found to be orbited by at least two planetary-mass objects, a 2.23~$R_\oplus$ super-Earth at an orbital period of 9.55~days and a {warm} 21-day period \mbox{3.95 ~$R_\oplus$} Neptune \citep[][hereafter C17 and R17]{Crossfield17,Rodriguez17}. 

The very nature of {warm} transiting planets such as HD106315c, namely their long orbital periods and thus rare transits events, 
poses a major limitation to their efficient further study because predicted ephemerides are uncertain. 
In the case of HD106315c\footnote{$P=21.0576^{+0.0020}_{-0.0019}$, $T_0=2457611.131\pm0.0012$ \citepalias{Crossfield17}}, the 3 $\sigma$ timing uncertainty for possible {early} 
JWST observations in mid-2019 amounts to 7.1~hours, making observations inefficient and challenging to schedule.

In this letter, we present ground-based transit observations HD106315c, and illustrate how flexible ground-based follow-up can resolve this issue for a large fraction of transiting planets expected from \emph{TESS}.

\section{Observations, data reduction, and analysis}
\subsection{Transit observations of HD106315~c}

\begin{table*}[ht]
\centering
\caption{\label{tab:obs}EulerCam observation details of 
HD106315. \tablefoottext{a}{at night start}}
\begin{tabular}{cccccccc} \hline \hline
Date\tablefootmark{a} & AM range & PSF FWHM [arcsec] & Exposure time [s] & baseline model & 5-min RMS [ppm] & $\beta_{w}$ & $\beta_{r}$ \\
\hline
2017-Mar-08 & 2.70 -- 1.13 -- 1.35 & 6.11 -- 7.76 & 30 & $t^2 max^1$ & 716 & 1.27 & 1.22  \\   
2017-Mar-29 & 1.66 -- 1.13 -- 2.60 & 7.08 -- 7.89 & 90 & $t^2 max^1$ & 713 & 1.40 & 1.31 \\  
\hline
\end{tabular}
\end{table*}

We observed HD106315 during two transits of planet c with EulerCam at the 1.2~m Euler-Swiss telescope at the ESO La Silla site. 
Both observations were carried out using an I-Cousins filter, applying a substantial telescope defocus to allow for an improved observation efficiency. 
The main properties of the observations are summarized in Table \ref{tab:obs} (see \citealt{Lendl12} for details on EulerCam and the data reduction procedures used to obtain the relative photometry). 

\subsection{MCMC analysis}

\begin{table}
\centering
\caption{\label{tab:res}Fitted and inferred parameters of HD106315c. \newline
\tablefoottext{a}{using stellar parameters of \citetalias{Crossfield17}.}}
\begin{tabular}{p{3.5cm}ll} \hline \hline
Parameter & Value & note \\
\hline
Radius Ratio, $R_p/R_\ast$ & $0.0315 \pm 0.0041$ & MCMC \\
Impact parameter, b & $0.43 \pm 0.30$ & MCMC \\
Transit duration, $T_{14}$ [d] & $0.1824_{-0.0071}^{+0.0084}$ & MCMC \\
First Euler mid-transit time [BJD-2450000] &  $7821.7030_{-0.0016}^{+0.0020}$ & MCMC \\
Second Euler mid-transit time  [BJD-2450000]& $7842.75341_{-0.0023}^{+0.0027} $ & MCMC \\
Transit midpoint, $T_0$ [BJD-2450000] & $7611.1313 \pm 0.0045$ & least-square \\
Period, P [d] & $21.05683 \pm 0.00053$ & least-square \\
First quadratic LD coefficient, $u_1$ & 0.49 & fixed \\
Second quadratic LD coefficient, $u_2$ & 0.32 & fixed  \\
Transit depth, dF & $992 \pm 258$~ppm & inferred \\
Planetary radius, $R_p$ & $4.05\pm 0.65$ & inferred\tablefootmark{a} \\
\hline
\end{tabular}
\end{table}

We used a Markov chain Monte Carlo approach to perform a joint analysis of both EulerCam observations. Our code combines the transit light curve model of \citet{Mandel02} 
with the publicly available differential evolution MCMC engine MCcubed \citep{cubillos16}. Throughout the analysis the following parameters were set as \textit{jump} 
(i.e., fitted) parameters: the relative planet-to-star radius ratio $R_p/R_\ast$, the timing of mid-transit $T_0$, the impact parameter $b$, the transit duration $T_{14}$, 
and the planetary orbital period $P$. To account for the a priori knowledge of these properties from the \emph{K2} data, we placed wide Gaussian priors on $T_0$, $b$, $T_{14}$, 
and $P$, centered on the values presented by \citetalias{Crossfield17} and with a width of three times their quoted 1 $\sigma$ errors for $b$ and $P$, and five times the quoted 
1 $\sigma$ uncertainties for $T_0$ and $T_{14}$. No prior was applied to $R_p/R_\ast$. 
Following the procedure established by, e.g., \citet{Gillon10a}, we tested a range of photometric baseline models, which are found by least-squares minimization at each MCMC step. 
We assume a second-order time polynomial as the minimum baseline model to our data and accept more complicated models only if favored by significant BIC improvement \citep[e.g.,][]{Schwarz78}. 
We tested a range of photometric baseline models for each light curve, accounting for external parameters 
while always performing a combined fit. We find that our data are best represented by assuming, 
for both light curves, the transit model and a photometric baseline consisting of a quadratic time polynomial and a linear dependence on the maximum target count rate. The 
latter we attribute to substantial PSF variations incurred with EulerCam when defocusing the telescope. 
Limb darkening was set constant throughout the analysis as the shallowness of the transit paired with the precision attainable from the ground do not allow meaningful 
constrains on the in-transit light curve curvature. We inferred appropriate quadratic limb darkening parameters (see Table \ref{tab:res}) using the routines by \citet{Espinoza15} together 
with the efficiency and filter transmission of EulerCam in Ic band, assuming a 6290~K star and interpolating the limb-darkening parameters to a $\log{g}$ of 4.29 \citepalias{Crossfield17}. 
We rescaled our photometric errors to account for underestimated red and white noise by calculating the $\beta_{r}$ and $\beta_{w}$ factors \citep{Winn08,Gillon10a} as described in \citet{Lendl13}. 
This was done on an initial least-squares fit of the data and using timescales between 5 and 20 minutes. Throughout our analysis, we checked MCMC convergence via the Gelman-Rubin test \citep{Gelman92}.

\section{Transit detection of HD106315c with EulerCam}

\begin{figure}[h!]
\includegraphics[width=0.95\columnwidth]{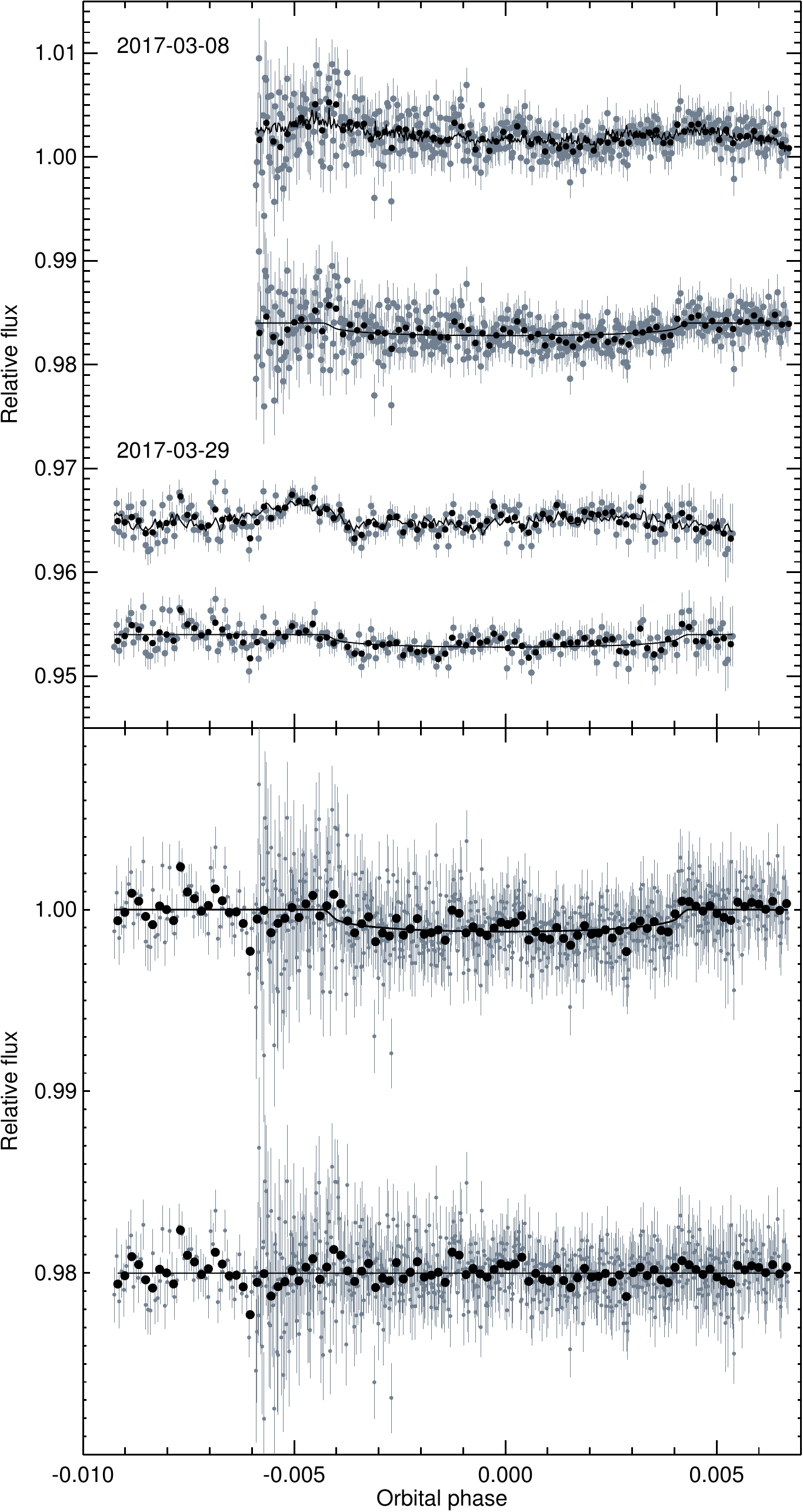}
\caption{\label{fig:lc}EulerCam observations of HD106315c. Top:  Raw light curves, together with their 
full (transit and baseline) photometric model. For each date, the uncorrected light curves are shown on top, and the 
light curves corrected for the photometric baseline model are shown below. Bottom:  Phase folded baseline-corrected data and residuals. Unbinned data are shown in gray, and the same data binned per 5 minutes are shown in black.}
\end{figure}

Based on the combined analysis of two EulerCam transit observations, we detect the transit of HD106315c with a depth of $992 \pm 258$~ppm, corresponding to $R_P/R_\ast =  0.0315 \pm 0.0041$. Our results are in excellent agreement with previously published values; however, we find a marginally shorter transit duration (1.3 and
1.2 $\sigma$ compared to \citetalias{Crossfield17} and \citetalias{Rodriguez17},
respectively). This could be a result of the low (30 min) cadence of the \emph{K2} data, resulting in a total of three data points obtained during ingress/egress. We present the individual transit light curves in the upper panel of Figure \ref{fig:lc}, together with their full photometric models. The corrected and combined light curve showcasing the detection is shown in the lower panel of Figure \ref{fig:lc}. 

We also inferred individual transit time measurements to aid in the search for transit time variations (TTVs), which may help to constrain planetary masses and system architecture. To this end, we separately analyzed each individual light curve while refitting photometric baseline models and constraining the transit shape to that found from our combined analysis using Gaussian priors. The resulting mid-transit times, listed in Table \ref{tab:res}, 
show a timing precision of $\sim$3~minutes, which is at the level of the predicted TTV variations in the HD106315 system in absence of significant eccentricities or additional planets \citepalias{Crossfield17}. 
Even so, a slope in radial velocity measurements indicates that the system likely contains additional planetary or stellar companions \citepalias{Crossfield17}. The transit timings presented here thus provide a starting point for more extensive TTV measurements constraining the mass and period of the exterior body.

Based on these transit timings, and the epoch derived from the \emph{K2} data by \citepalias{Crossfield17}, we redetermine the planetary period via least-squares minimization. 
Doing so, we refine the planetary ephemeris, and near-quadruple the precision on the planetary period, significantly improving timing predictions for subsequent follow-up observations.

\section{Ground-based follow-up of transiting planets detected from space}
The detection of a relatively long-period (21 days), Neptune-sized planet from the ground with a one-meter class telescope opens exciting perspectives for the photometric follow-up of transit surveys. \emph{TESS} is expected to be launched in 2018 and will survey the whole sky in search of new transiting planets. \citet{Sullivan15} have simulated this mission yield and estimate that \emph{TESS} could detect $\sim$1\,700 transiting planets from $\sim$200\,000 pre-selected stars. \emph{TESS} will mainly detect planets smaller than $\sim$4~R$_\oplus$. Across most of the sky, \emph{TESS} will only detect (by observing two transits) {and} confirm (by observing at least three transits) planets with orbital periods shorter than 13.7 or 9.1 days, respectively. This is due to the mission's observing strategy of surveying the sky in 26 $24\degr \times 24\degr$ fields, each of them observed for a duration of 27.4~days ($27.4 / 2 = 13.7$; $27.4 / 3 = 9.1$). Fields overlap at the ecliptic poles, which receive nearly continuous coverage for 355 days. However, only $\sim30\%$ of \emph{TESS} planets are found within $20\degr$ of the ecliptic poles after \citet{Sullivan15}; hence, the majority of Neptune-sized planet candidates at periods greater than 9.1~days will have less than three observed transits. Following up these candidates and obtaining a third transit to confirm the period, refine ephemerides, and check for transit timing variations (TTVs) could be performed from the ground. In the following, we investigate how many of these long-period, Neptune-sized \emph{TESS} candidates could be confirmed with a one-meter telescope such as Euler.

\begin{figure}
\centering
\includegraphics[width=0.9\columnwidth]{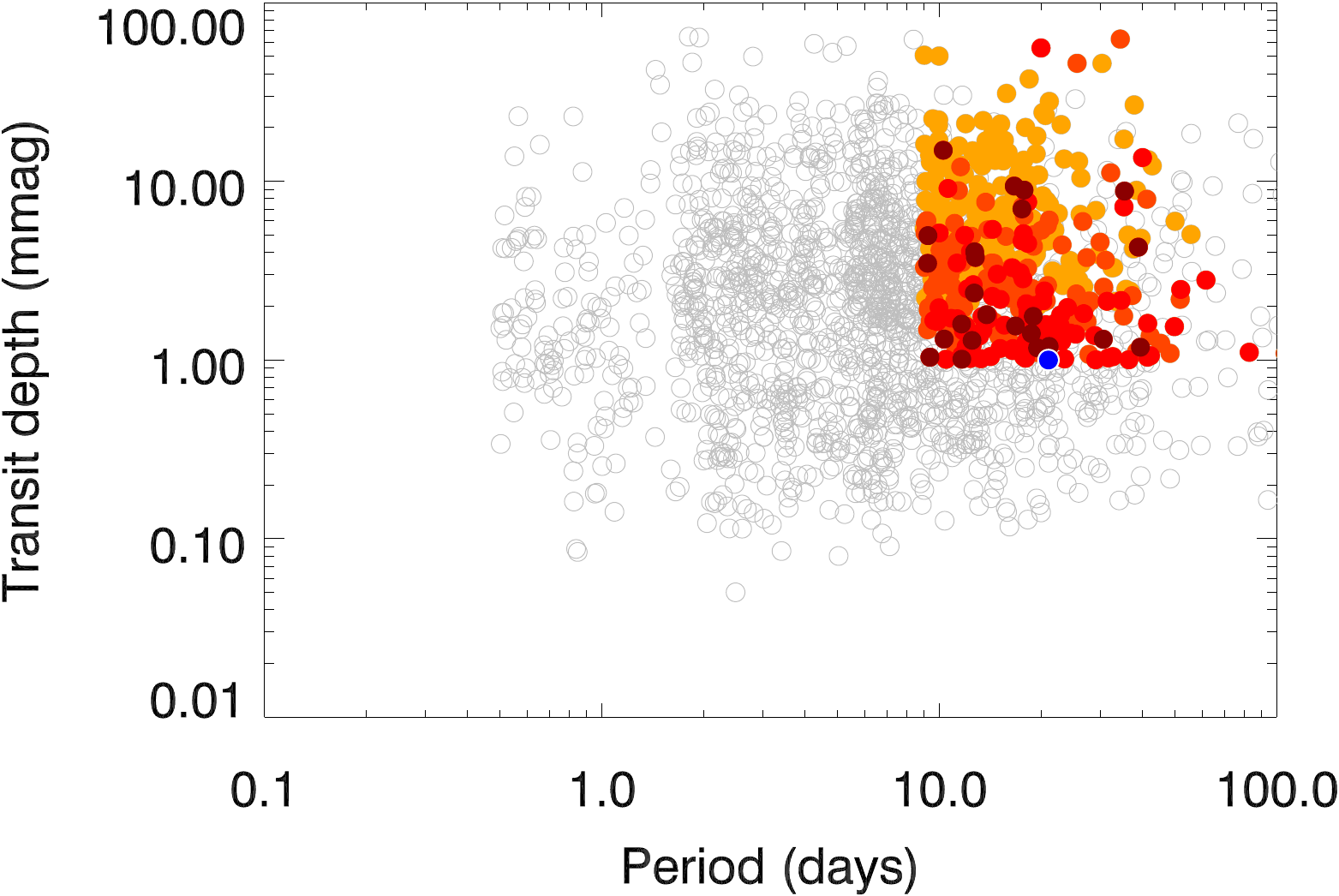}
\caption{\label{fig:TESSyield} Transit depth vs.\ revolution period for planets in the simulated yield of \emph{TESS} over the whole sky \citep[][gray circles]{Sullivan15}. Planet candidates with sizes $\leq 5$~R$_\oplus$, transit depths $\geq 1$~mmag, and periods $\geq 9.1$~days located $\geq 20\degr$ away from the ecliptic poles (i.e.,\ with $<3$ transits) are shown as orange dots. Those candidates are further distinguished based on their host star magnitude: stars with $V\leq 14$ (orange red), $V\leq 12$ (red), and $V\leq 10$ (dark red). HD~106315c is represented by the blue dot.}
\end{figure}

\begin{figure}
\centering
\includegraphics[width=0.9\columnwidth]{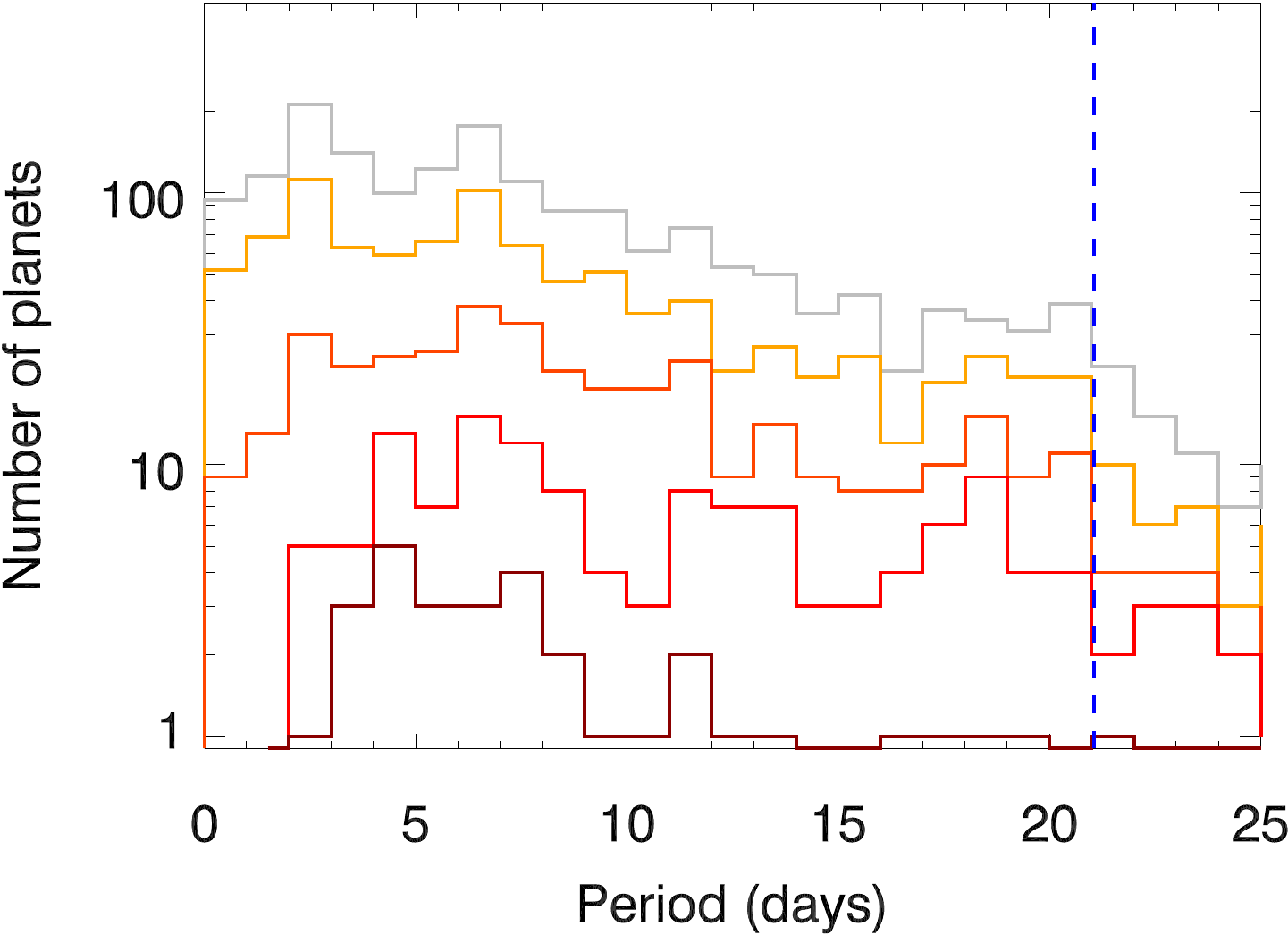}
\caption{\label{fig:TESShisto} Number of planet candidates ($R \leq 5$~R$_\oplus$; $\Delta m \geq 1$~mmag) as a function of revolution period for the whole \emph{TESS} sample (gray) and candidates located $\geq 20\degr$ away from ecliptic poles (orange) with host stars brighter than $V = 14$ (orange red), $V = 12$ (red), and $V = 10$ (dark red). The period of HD~106315c is indicated by the blue vertical dashed line.}
\end{figure}

We make use of the simulated \emph{TESS} yield from \citet{Sullivan15} to estimate how many planet candidates smaller\footnote{This limit is set to remove the (few) gas giants expected in the \emph{TESS} yield.} than 5~R$_\oplus$ on orbits longer than 9.1~days produce transits deeper than 1~mmag that are potentially detectable with Euler. We exclude candidates within $20\degr$ of the ecliptic poles to retain only those candidates with less than three observed transits. Figure~\ref{fig:TESSyield} shows the transit depth as a function of revolution period for the whole yield. The long-period Neptune-sized candidates interesting for ground-based follow-up are colored depending on their host star apparent $V$ magnitude. Figure~\ref{fig:TESShisto} presents these results as histograms to better quantify the number of planet candidates based on their host star magnitudes. 

We find 548 Neptune-sized planet candidates with periods longer than 9.1~days within the whole \emph{TESS} simulated yield; 306 of them (56\%) are located $20\degr$ away from the ecliptic poles and thus cannot benefit from an extended temporal coverage: more than half of the long-period Neptune-sized planets from \emph{TESS} will require photometric follow-up in search of a third transit. Among these 306 candidates, there are 162 transit stars with $V\leq 14$, 74 transit stars with $V\leq 12$, and 12 transit stars brighter than $V=10$. \footnote{In the $I$ ($J$) band, there are 293 (306), 177 (266), and 40 (90) long-period Neptune-sized candidates transiting stars brighter than $I (J) = 14$, 12, and 10, respectively.}

Recently, \citet{Fulton17} have shown revised occurrence rates for small planets. We investigate what effect, if any, this may have on the original estimates of \citet{Sullivan15} by comparing their occurrence rates for stars with $T_{\rm eff}$ between 4700 and 6500~K to those of \citet{Fulton17}. We do this by re-binning the occurrence rate as a function of radius given by \citet{Fulton17} into the radius bins used by \citet{Sullivan15} to generate planet radii. We then up-scale the final \emph{TESS} yield by the ratios of the occurrence rates. By doing so we estimate \emph{TESS} could yield $700\pm175$ extra planets during its two-year primary mission. Most of these would be sub-Neptunes with radii between 2 and 4~R$_\oplus$. 

As can be seen in Figure~\ref{fig:TESSyield}, shallower transits are found preferentially around bright stars. 
A dozen of HD~106315c-like candidates with similar magnitudes ($V\leq 10$) are expected from \emph{TESS}. These objects will be prime targets for detailed atmospheric studies and thus high priority objects for fast photometric and spectroscopic follow-up. Nonetheless, the fainter ($10\leq V \leq 14$) magnitude range also contains a promising pool of objects, in particular with regard to their smaller overall host star radii. As the fractional size ratio between the planet's atmosphere and the host star is enhanced for small stars, these systems possess more readily observable atmospheres. At the same time, their planets are cooler than planets orbiting solar-type stars with the same orbital period, providing us with a window on less irradiated planetary atmospheres. Observationally, these slightly fainter objects are also excellent targets for ground-based follow-up as they typically have many nearby references stars of similar brightness, while wide-passband filters\footnote{EulerCam is equipped with the 515 --880 nm high-throughput NGTS filter} assure that observations are not limited by photon noise down to V$\approx$14. Limited availability of nearby reference stars can be an issue for bright candidates. We find that all predicted $V < 10$ TESS targets have at least one $\Delta V < 2$ star and 67 \% of them have at least three $\Delta V < 1$ stars within 15'. This warrants that most of the 162 simulated candidates identified could be effectively followed up with high-precision photometry from the ground. 

For a handful of targets with $13<V<14$, predicted TESS precisions are comparable or inferior to those obtained here. For these objects, ground-based observations can help to improve the precision on the planetary radius. Furthermore, the angular resolution of Euler (seeing limited, $\sim$~0.9 arcsec) is greatly superior to that of TESS (21.1 arcsec per pixel), allowing us to identify false positives due to blended eclipsing binaries.

\section{Conclusions}
We present the detection of the \mbox{$\la 1$~mmag} transit of HD~106315~c, a 21-day {warm} Neptune using a ground-based 1 m class telescope. Pinning down the period of the object, which had been discovered by \emph{K2} based on only two individual transits, we show the potential of ground-based follow-up of small transiting planets discovered from space. Based on the predicted \emph{TESS} yield, and using the attained precision as a {conservative} estimate, we envision that ground-based observations with small telescopes can provide fast and affordable means of obtaining precise ephemerides for a large sample of Neptune-sized planets discovered by \emph{TESS}. This effort will critically ease the weight on other high-performance photometric telescopes, for example\ CHEOPS \citep{Broeg13}, while enabling follow-up studies of {warm} Neptune-sized planets. It could be tackled by several ground-based telescopes already in operation (LCOGT, \citealp{Shporer11}, TRAPPIST, \citealp{Gillon11a, Jehin11}) or coming online in the next years (SaintEX, SPECULOOS). This study demonstrates the feasibility of this approach.

\begin{acknowledgements}
We would like to thank an anonymous referee for the fast and constructive comments that have improved the quality of this work and I.Ribas for facilitating access to the TESS simulated yield. This work has been carried out within the framework of the NCCR PlanetS supported by the Swiss National Science Foundation. D.E. acknowledges funding from the European Research Council (ERC) under the European Union’s Horizon 2020 research and innovation program (grant agreement No 724427).
\end{acknowledgements}

\bibliographystyle{aa}
\bibliography{bbl}

\end{document}